\documentclass[aps,prresearch,twocolumn]{revtex4-2} 

\usepackage[]{graphicx}
\usepackage[dvipsnames]{xcolor}
\usepackage[colorlinks=true,citecolor=NavyBlue,linkcolor=RubineRed,urlcolor=Cerulean,pdfencoding=auto]{hyperref}
\usepackage[]{amsmath}
\usepackage[]{amssymb}
\usepackage{tabstackengine}

\begin{document}
\title{Dynamic  sweet spot of  driven flopping-mode spin qubits in planar quantum dots}

\author{Yaser Hajati}
\email[]{yaser.hajati@uni-konstanz.de}
\affiliation{Department of Physics, University of Konstanz, D-78457 Konstanz, Germany}

\author{Guido Burkard}
\email[]{guido.burkard@uni-konstanz.de}
\affiliation{Department of Physics, University of Konstanz, D-78457 Konstanz, Germany}

\begin{abstract}
  Semiconductor quantum dots with confined electron or hole spins show promise for quantum information processing as they allow for efficient electric field-driven qubit manipulation. However, their susceptibility to electric noise poses a challenge that may hinder the effectiveness of these qubits. Here, we explore the impact of electric noise on a planar double quantum dot (DQD) spin qubit under the influence of AC gates applied to the dot levels, focusing on the flopping-mode spin qubit with spin-orbit interaction. We employ a rotating wave approximation within a time-dependent effective Hamiltonian to derive analytic expressions for the Rabi frequency of spin qubit oscillations with a single electron or hole in a DQD. We find that driving the qubit off-resonantly effectively mitigates the influence of charge noise, leading to a manifestation of a dynamic sweet spot. The proposed mode of operation notably improves the fidelity of quantum gates, particularly within specific ranges of drive parameters and detuning during qubit manipulation. Furthermore, our study unveils the potential of inducing a second-order dynamic sweet spot, a phenomenon tunable by drive and DQD parameters. Understanding the importance of driving qubits off-resonantly is essential for developing high-coherence planar DQD spin qubits, both for electrons in silicon and holes in germanium.
\end{abstract}

\maketitle

\section{Introduction}


The pursuit of a fault-tolerant quantum computer capable of solving complex problems relies on integrating highly coherent qubits \cite{Terhal,reiher}. Quantum dots (QDs), fundamental to spin qubits, offer promise for advancing large-scale quantum processors due to their compact size and sustained coherence \cite{loss}. Recent progress in single-qubit and two-qubit gate operations, alongside high-precision readout, underscores advancements in quantum dot-based quantum computing \cite{maurand, meno1, harvey, zajac,guido}. Electric dipole spin resonance (EDSR), incorporating spin-orbit interaction (SOI), hyperfine interaction, and g-factor modulation, provides a robust method for electrically manipulating electron spin states \cite{nowack, vitaly, lirad, kato, watson, Huang,maurand2}. With its reliance on purely electric fields, EDSR stands out as a well-established technique within the QD domain, offering advantages over electron spin resonance (ESR), notably in energy efficiency and ease of localization and selectivity \cite{kawakami, yoneda, franck, nowack}.

Notably, the flopping-mode concept in double quantum dot (DQD) structures has driven the progress towards achieving strong spin-photon coupling \cite{monica, monica2, croot, cubay, sami, mi}. This mode of operation involves the delocalization of the charge of one electron between both dots, significantly enhancing the coupling strength to the driving or cavity electric field and facilitating efficient information transfer between electron-spin qubits and microwave photons. While silicon and carbon DQDs have been extensively studied for their potential as flopping-mode spin qubits, relying on synthetic SOI induced by a magnetic field gradient, heavy holes (HHs) in germanium are found to naturally form a flopping-mode qubit without the need for synthetic SOI \cite{philip}. This inherent characteristic of the Ge flopping-mode spin qubit not only simplifies its realization in the laboratory but also renders it less susceptible to errors arising from imperfections in dot engineering compared to its conduction band counterpart \cite{monica}.

 The hybridization of spin and charge degrees of freedom also leads to a certain impact of charge noise on the flopping-mode spin qubit. This effect can be counteracted by working on so-called charge-noise sweet spots when the qubit is idle, effectively extending decoherence times \cite{monica,croot}. 
 In recent years, various studies \cite{pic,yang,mun,val,nico,hua,Frees,teske} have extended this concept to the protection of the qubit during AC gates at so-called dynamical sweet spots, which can improve gate fidelities. In dynamic scenarios, such as employing an AC gate to drive a qubit and induce Rabi oscillations, the system dynamically adjusts control parameters such as the Rabi frequency in real-time to counteract environmental fluctuations, optimizing qubit performance throughout the computation. In this study, we investigate the impact of charge noise on a planar DQD spin qubit subject to AC gates, relevant for both electrons in silicon (Si) and holes in germanium (Ge), focusing on the flopping-mode spin qubit with SOI. We employ a rotating wave approximation (RWA) within a time-dependent Schrieffer-Wolff effective Hamiltonian to derive analytic expressions for the Rabi frequency of DQD qubit oscillations for a single electron or hole spin. Our results emphasize the effectiveness of strategically driving the qubit off-resonantly in mitigating the impact of charge noise, resulting in the emergence of a dynamic sweet spot. This modulation significantly enhances the fidelity of quantum gates, particularly within defined ranges of drive parameters and detuning during qubit manipulation. Understanding the significance of driving qubits off-resonantly is pivotal in designing high-coherence planar DQD spin qubits in silicon and germanium.

%

\begin{figure}[!ht]
\centering
\includegraphics[width=1\linewidth]{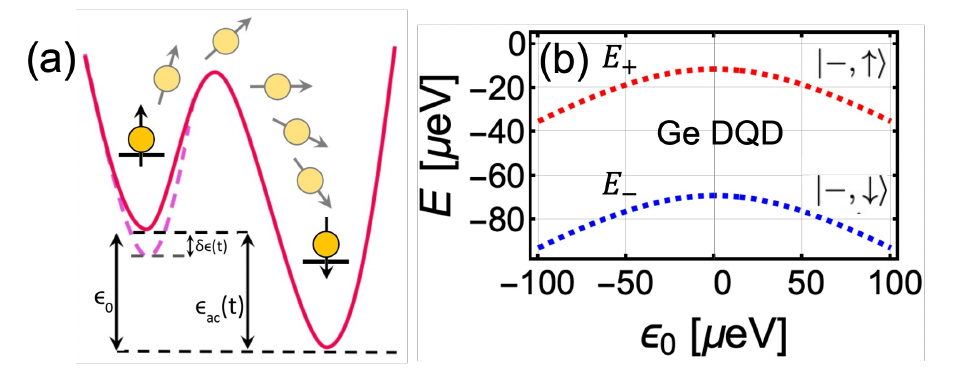} 
\caption{\label{fig:level1}
  (a) Schematic illustration of a flopping-mode spin qubit, showing the delocalization of the associated charge carrier (electron or hole) between two dots through time-dependent driving in the presence of SOI. Here, $\epsilon_{\text{ac}}(t)$ and $\delta\epsilon(t)$ are AC control of the detuning parameter and detuning fluctuations caused by charge noise, respectively. (b) Energy levels \( E_{+,-} \) of a Ge DQD (see  Eq.~\eqref{eq:b4} in Appendix~\eqref{eq:B}) as a function of detuning $\epsilon_0$ for $\eta = 0.8   $ $\mu$eV, $\chi = 0.1$ $ \mu$eV, $E_z = 57.8$  $\mu$eV, and $t_c = 40$ $ \mu$eV. The state vectors $|-, \uparrow \rangle$ and $|-, \downarrow \rangle$ represent the bonding states with opposite spin orientations. 
  }
\end{figure}

%

The remainder of this paper is structured as follows. Section~\ref{sec:stiment} introduces the flopping-mode spin qubit for Ge DQD with natural SOI and Si DQD with synthetic SOI. In Sec.~\ref{sec:stimmodel}, we apply a RWA within a time-dependent Schrieffer-Wolff transformed effective Hamiltonian to derive analytical expressions for the Rabi frequency of single electron or hole DQD qubit oscillations. The behavior of the flopping-mode spin qubit for Ge and Si DQDs is investigated under both on-resonance and off-resonance driving in Sec.~\ref{sec:stimresults}. Finally, our results are summarized, and conclusions are drawn in Sec.~\ref{sec:stimconc}.

\section{Flopping-mode DQD spin qubit }\label{sec:stiment}
We model the flopping-mode spin qubit as a single electron (or hole) in a DQD, delocalized between two dots in a magnetic field, in the presence of synthetic (natural) SOI, as illustrated in Fig.~\ref{fig:level1}(a). We now present the full theoretical model used in this work for a single electron in a Si DQD or a single hole in a Ge DQD hybrid spin qubit. The Hamiltonian is given by 
\begin{align}
 \label{eq:Hsys}
  H(t)=H_{\rm DQD}+H_{\rm ac}(t)+H_{\rm SO}, 
 \end{align} 
where the three components of the Hamiltonian are described below. 
The double quantum dot Hamiltonian $H_{\rm DQD}$ for a single electron (or hole) in a DQD in a homogeneous magnetic field is given by $H_{\rm DQD} = (\epsilon_{0}/2)\tau_{z} + t_c\tau_{x} + (E_z/2)\sigma_{z}$, where $\epsilon_{0}$ is the energy detuning between the two dots, $t_c$ is the interdot tunnel coupling, $\tau_i$ and $\sigma_z$ denote the Pauli operators in the charge (L,R) and spin ($\uparrow,\downarrow$) subspaces. Here, the Zeeman energy associated with the applied homogeneous out-of-plane magnetic field $B_z$ is defined as $E_z=g\mu_B B_z$, with $g = 2(10)$  the g-factor for an electron in silicon (hole in germanium), and $\mu_B$ is the Bohr magneton \cite{frey,jirovec,philip}. In a symmetric DQD with zero energy-level detuning ($\epsilon_{0} = 0$) and in the absence of any magnetic field, the electron or hole in the DQD forms bonding and antibonding charge states $|\mp\rangle =(|R\rangle \mp|L\rangle )/\sqrt{2}$, separated by an energy of $2|t_c|$\cite{stock,bru,kim}. In the highly detuned regime ($|\epsilon_0| \gg |t_c|$), the electron or hole is strongly localized in either the left $|L\rangle $ or right $|R\rangle $ dot. The dominant bonding-antibonding energy difference, $2|t_{c}|$, makes the charge highly susceptible to oscillating electric fields. 
\\For the AC drive, $H_{\rm ac}$, we consider a time-dependent detuning through an AC gate voltage as $\epsilon_{\rm ac}(t)=(A/2) \tau_{z} \cos(\omega_{d} t)$, see Fig.~\ref{fig:level1}(a), where A and $\omega_d$ denote the drive amplitude and frequency, respectively. 
    Finally, $H_{\rm SO}$ describes the (synthetic) spin-orbit interaction. For a single electron in a silicon DQD in the presence of an inhomogeneous magnetic field from a micromagnet combined with a homogeneous external magnetic field, the synthetic SOI is modeled using the Hamiltonian $H_{\rm SO}=b_{x}\tau_{z}\sigma_{x} + b_{z}\tau_{z}\sigma_{z} $, where $b_{x}$ and $b_{z}$ correspond to transverse and longitudinal magnetic gradient fields   which can hybridize the spin and charge degrees of freedom of an electron in a silicon DQD. Specifically, a micromagnet can be tailored to induce a longitudinal magnetic field gradient between the left and right QDs, aiming to achieve distinct spin resonance frequencies based on the position of the electron. The synthetic SOI is induced by the magnetic field gradient, effectively hybridizing states with opposing spin directions into bonding and antibonding configurations. This interaction is characterized by the two spin-orbit mixing angles, which can be expressed as $\phi_{\pm}=\arctan[\mu_{B}B_{z}b_{x}/(2t_{c}{\pm} E_{z})]$, where $\phi_{\pm}$ ranges within $[0, \pi]$. We also investigate a semiconductor heterostructure with a germanium quantum well, confining HHs robustly along the out-of-plane (z) axis. Incorporating cubic Rashba SOI \cite{Bul1,Bul2} and considering higher-orbit effects, specifically spin-flip tunneling SOI $\eta$, and intradot spin-flip SOI $\chi$, we model the SOI of a single hole in a Ge DQD using the Hamiltonian \cite{philip} as $H_{\rm SO}= \chi\tau_{z}\sigma_{x}+ \eta\tau_{y}\sigma_{y}$, where the parameter $\chi$ plays a role analogous to that of $b_{x}$ for electrons in silicon. It is found that the intradot spin-flip SOI $\chi$ in germanium serves as an intrinsic magnetic gradient field  with a strength of 2$\chi$ \cite{philip}. Although this term emerges from a perturbative approach in the SOI, its magnitude can be substantial, independent of the overlap between the left and right dot states. For a magnetic field strength $B=1 {\rm T}$, the value of 2$\chi$ is approximately 0.6 $\mu$eV, considering a HH-LH splitting of 100 meV and a Rashba coefficient $\alpha_{R} \langle E_{z} \rangle$ of $10^{-11}$ eVm. This magnitude is comparable to the magnetic field gradients applied in electronic systems, where $g \mu_B b_x \approx 1.6 \, \mu\text{eV}$ \cite{monica3,mi}. It is crucial to highlight that the synthetic SOI terms, \( b_x \) and \( b_z \), in Si DQDs and the natural SOI terms, \(\eta\) and \(\chi\), in Ge DQDs result in coupling between the spin and charge subspaces. To achieve spin rotation and Rabi oscillation, only one of these terms is necessary for each respective material.


\section{Time-dependent Schrieffer-Wolff transformation and rotating wave approximation}\label{sec:stimmodel}
Initially, weonalize of the Hamiltonian in Eq.~\eqref{eq:Hsys} in the absence of any Zeeman field, SOI, and driving. This diagonalization results in the emergence of bonding states with lower energy and antibonding states with higher energy within the DQD system as $E_{\pm} = \pm \sqrt{t_c^2 + \epsilon_0^2/4}$. To comprehend the interplay between these states, we extend our analysis to transform the total Hamiltonian H(t) in the presence of the Zeeman field, SOI, and periodic driving into the bonding-antibonding basis,
%
\begin{widetext}
\begin{equation}
\begingroup 
\setlength\arraycolsep{-4pt}
H(t) = \begin{pmatrix}
    K_{+} + \frac{1}{2}\left(A_{c} \cos(\omega_{d}t) + b_{zc}\right) & \rho_{c} & -\frac{1}{2}\left(b_{zs} + A_{s} \cos(\omega_{d}t)\right) & -\eta - \rho_{s} \\
    \rho_{c} & K_{-} + \frac{1}{2}\left(A_{c} \cos(\omega_{d}t) - b_{zc}\right) & \eta - \rho_{s} & \frac{1}{2}\left(b_{zs} - A_{s} \cos(\omega_{d}t)\right) \\
    -\frac{1}{2}\left(b_{zs} + A_{s} \cos(\omega_{d}t)\right) & \eta - \rho_{s} & -K_{-} - \frac{1}{2}\left(A_{c} \cos(\omega_{d}t) + b_{zc}\right) & -\rho_{c} \\
    -\eta - \rho_{s} & \frac{1}{2}\left(b_{zs} - A_{s} \cos(\omega_{d}t)\right) & -\rho_{c} & -K_{+} - \frac{1}{2}\left(A_{c} \cos(\omega_{d}t) - b_{zc}\right) \\
\end{pmatrix},
\endgroup
\label{eq:Ham}
\end{equation}
\end{widetext}
where we define $\rho_c = \rho \cos(\phi)$, $\rho_s = \rho \sin(\phi)$, with $\rho = \chi(b_x/2)$ for a hole (electron) in a germanium (silicon) DQD, $b_{zc} = b_z \cos(\phi)$, $b_{zs} = b_z \sin(\phi)$, $A_c = A \cos(\phi)$, $A_s = A \sin(\phi)$ with the DQD mixing angle $\phi=\arctan(2t_{c}/\epsilon_{0}) $ and $K_{\pm} = \sqrt{t_c^2 + \epsilon_0^2/4} \pm E_z/2$. Also, for a hole (electron) in a germanium (silicon) DQD,  $b_z = 0$ ($\eta = 0$). Note also that in our calculations for a Si DQD, we have neglected $b_{z}$ except in Fig.~\ref{fig:level6} where \( b_z = 4 \, \mu\text{eV} \). We employ a second-order time-dependent Schrieffer-Wolff transformation (TDSWT) to derive a time-dependent effective Hamiltonian for a single hole or electron in a DQD qubit \cite{Romi}. Hence, the Hamiltonian in Eq.~\eqref{eq:Ham} following the application of the TDSWT yields a block-diagonal structure,
\begin{equation}
H_{\rm TDSWT}(t) = \begin{bmatrix}
    H_{\text{Bonding}} & 0_{2\times2} \\
    0_{2\times2} & H_{\text{AntiBonding}}
\end{bmatrix},
\label{eq:TDSWT}
\end{equation}
where $H_{\text{Bonding}} $ and $H_{\text{AntiBonding}}$ represent 2×2 Hamiltonians corresponding to the bonding and antibonding states for both Ge and Si DQDs. For a detailed derivation, refer to Appendix~\ref{eq:A}. As the qubit resides in the bonding states, we focus solely on the Hamiltonian for the bonding state \cite{philip}. Employing the WA, we derive $H_{\text{RWA}}$ in the rotated frame as
\begin{equation}
H_{\text{RWA}} = UH_{\text{Bonding}}(t)U^\dagger + i \dot{U} U^\dagger,
\label{eq:RWA1}
\end{equation}
where $U = \text{diag}[e^{-i\omega_d t/2}, e^{i\omega_d t/2}]$. Consequently, we can determine the corresponding $H_{\text{RWA}}$ and the frequency of Rabi oscillation $\Omega$ within the slightly off-resonant regime (i.e., \(\omega_d \neq \omega_0\)), as detailed in Appendix~\ref{eq:B},
\begin{align}
H_{\text{RWA}} &= \Delta\omega \sigma_z + \Omega \sigma_x, \label{eq:RWA} \\
\Omega &= \sqrt{\Omega_0^2 + \Delta\omega^2}, \label{eq:R}
\end{align}
where $\Delta\omega = \omega_d - \omega_0$ is the detuning between driving and resonance frequencies. Here, the Rabi frequency reduces to $\Omega=\Omega_0$ for on-resonance driving (i.e., $\omega_d = \omega_0$), where $\omega_0$ denotes the energy level splitting (or resonance frequency) of the spin qubit. The energy levels of the bonding states, $H_{\text{Bonding}}$ in Eq.~\eqref{eq:TDSWT}, following the TDSWT treatment and RWA, are depicted as a function of detuning $\epsilon_0$ for Ge DQD in Fig.~\ref{fig:level1}(b) (see Eq.~\eqref{eq:b4} in Appendix~\eqref{eq:B}). Significantly, these levels exhibit a favorable agreement with the energy levels of bonding states in Ge DQDs in the absence of TDSWT and RWA, as reported in Ref.~\cite{philip}. Note that for the bonding states of Si DQDs, which are not shown here, we see a good agreement with Ref.~\cite{monica}. We find that TDSWT works well, effectively distinguishing between the bonding and antibonding states in the regime where $2t_c > B_z$, with weak coupling between these states.
\begin{figure*}[ht!]
\centering
\includegraphics[width=0.8\linewidth]{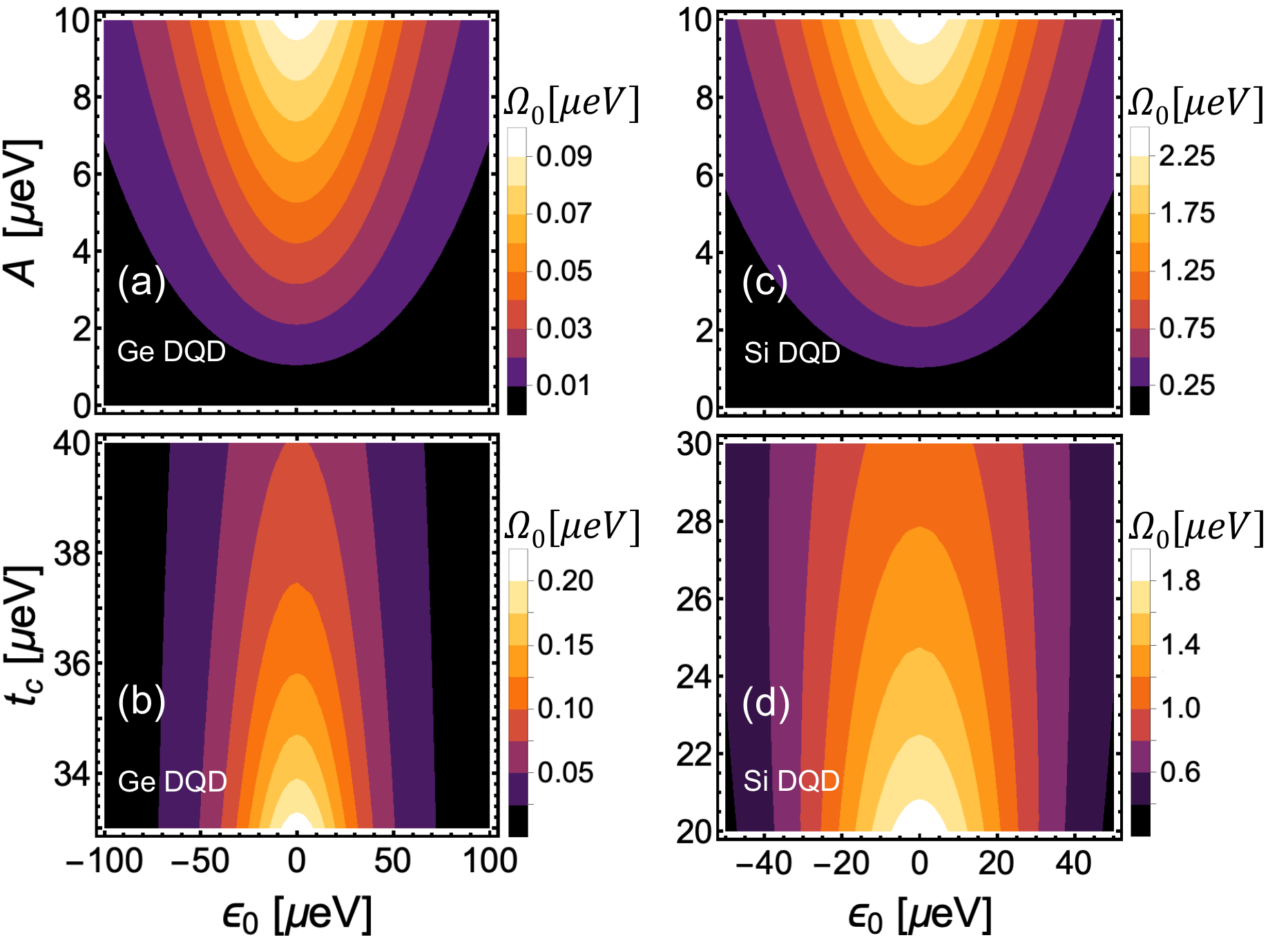} 
\caption{\label{fig:level2}
   (a) and (c) Rabi frequency $\Omega_0$ as a function of drive amplitude \(A\) and detuning $\epsilon_{0}$  for Ge and Si DQDs under on-resonance driving (i.e., \(\omega_d = \omega_0\)), respectively. (b) and (d) Rabi frequency $\Omega_0$ as a function of tunnelling coupling $t_{c}$ and detuning $\epsilon_{0}$ for Ge and Si DQDs under on-resonance driving, respectively. In the Ge DQD case, with \( \eta = 0.8 \, \mu\text{eV} \), \( \chi = 0.1 \, \mu\text{eV} \), \( E_z = 57.8 \, \mu\text{eV} \), (a) corresponds to \( t_c = 40 \, \mu\text{eV} \), and (b) to \( A = 8 \, \mu\text{eV} \). For the Si DQD case, with \( b_x = 15 \, \mu\text{eV} \), \( b_z = 0 \), \( E_z = 24 \, \mu\text{eV} \), (c) represents \( t_c = 20 \, \mu\text{eV} \), and (d) \( A = 8 \, \mu\text{eV} \).
}
\end{figure*}
In semiconductor DQD architectures, the pervasive presence of charge noise introduces fluctuations in the electrostatic potentials between the left and right QDs. Consequently, this charge noise predominantly manifests in the DQD system as variations in the detuning parameter $\epsilon_{0}$, transforming it to $\epsilon_{0}+\delta\epsilon_{0}$. The fluctuations in $\epsilon_{0}$ are modeled as quasistatic and Gaussian-distributed, where quasistatic implies that $\delta\epsilon_{0}$ remains constant within a single experimental run but varies across different runs. To account for this noise, we incorporate it into our calculations by convolving the relevant quantities with a Gaussian distribution. The assessment of the effectiveness of the dynamic sweet spot in the proposed DQD can be quantified through the average gate fidelity \cite{Hjor} given by,
\begin{equation}
F = \frac{d + |\text{Tr}(U_{\text{ideal}}^T U_{\text{actual}})|^2}{d(d+1)},
\label{eq:F}
\end{equation}
where \( d \) represents the dimension of the Hilbert space ($d=2$ for a single-qubit gate operation), \( U_{\text{ideal}} \) denotes the desired qubit operation, which in our scenario corresponds to a Rabi oscillation in the off-resonant regime, and \( U_{\text{actual}} \) represents the actual qubit operation in the off-resonant regime, incorporating undesired static charge noise where \( \epsilon_{0} \) is perturbed such that it turns into \( \epsilon_{0} + \delta\epsilon_{0} \).

\section{Results}\label{sec:stimresults}
In Fig.~\ref{fig:level2}, we show the Rabi frequency $\Omega_0$ as a function of detuning $\epsilon_0$, drive amplitude $A$, and tunneling coupling $t_{c}$ under  driving on-resonance (i.e., $\omega_d = \omega_0$) for both the Ge and Si DQDs qubit scenarios. Our observations in Fig.~\ref{fig:level2}(a) (Fig.~\ref{fig:level2}(c)) reveal an intriguing relationship: the Rabi frequency associated with a hole (electron) confined within a planar DQD, when driven resonantly, showcases an inverse correlation with detuning energy $\epsilon_{0}$ while concurrently displaying a positive correlation with drive amplitude $A$. Notably, due to the significantly larger synthetic SOI in Si ($b_x = 15 \, \mu\text{eV}$) compared to the natural SOI in Ge ($\eta = 0.8 \, \mu\text{eV}$ and $\chi = 0.1 \, \mu\text{eV}$), the Rabi frequency in Si DQD surpasses that in Ge DQD, reaching around $2.25 \, \mu\text{eV}$ at $A = 10 \, \mu\text{eV}$ and $\epsilon_0 = 0$, as depicted in Fig.~\ref{fig:level2}(c). In Fig.~\ref{fig:level2}(b) (Fig.~\ref{fig:level2}(d)), we have plotted the Rabi frequency $\Omega_0$ as a function of detuning $\epsilon_0$ for different tunneling coupling $t_{c}$ for a Ge DQD (Si DQD). We observe that increasing the tunneling coupling $t_{c}$ leads to a decrease in the Rabi frequency. It is important to avoid further decreasing the tunneling coupling $t_c$ to prevent nearing the resonance condition of the DQD where $g\mu_{B}B_z=\sqrt{\epsilon_0^2+t_{c}^2}$), such that the energies of the bonding and antibonding states align, leading to a sharp change in coupling. In such a scenario, the TDSWT breaks down.

\begin{figure*}[ht!]
\centering
\includegraphics[width=0.9\linewidth]{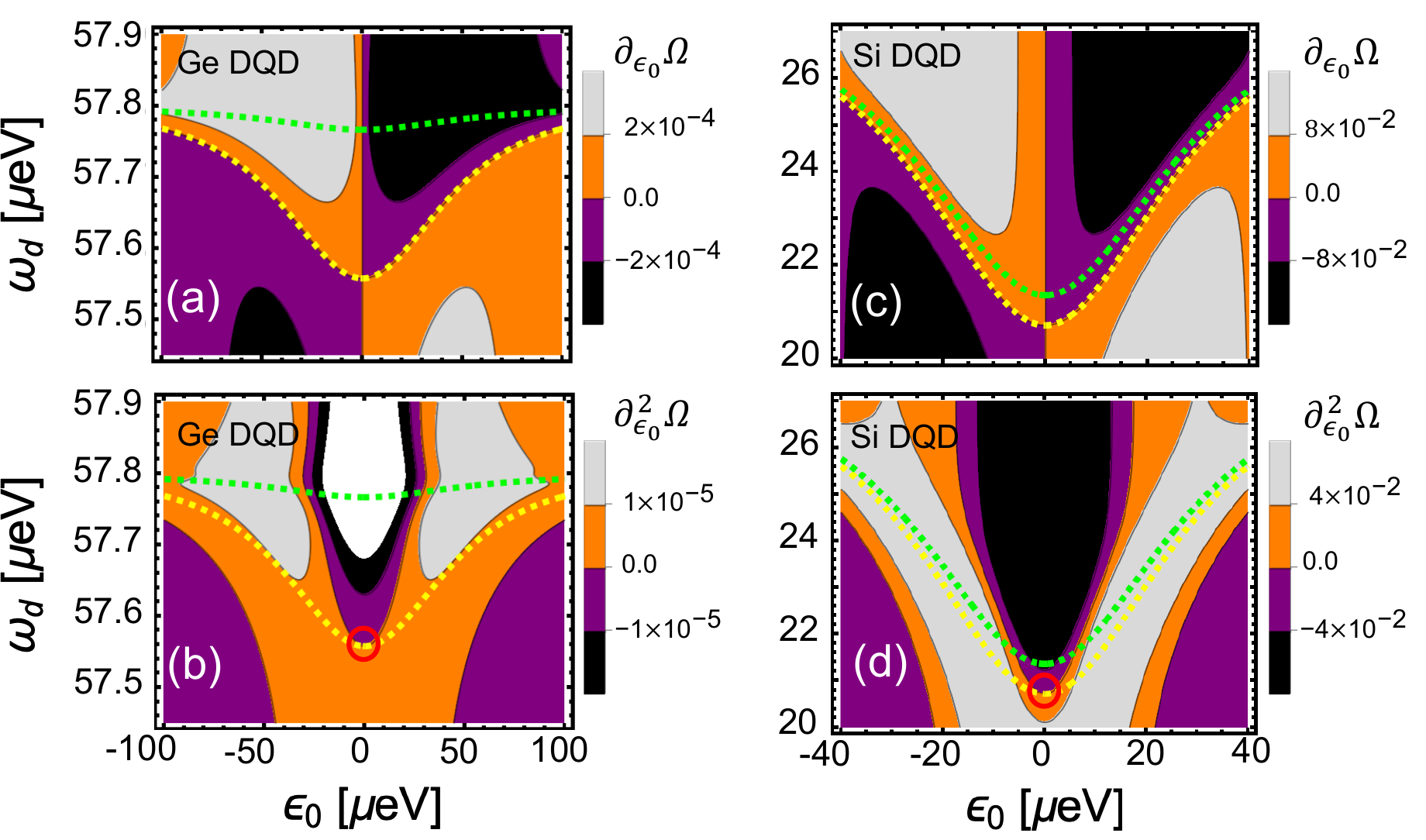}
\caption{The first and second derivatives of the Rabi frequency $\Omega$ as a function of drive frequency $\omega_d$ and detuning $\epsilon_0$ are depicted for a Ge DQD in (a) and (b), and for a Si DQD in (c) and (d), respectively. The yellow dashed line indicates the emergence of the first-order dynamic sweet spot (i.e., $\partial_{\epsilon_{0}}\Omega=0$), while the green dashed line represents the resonance frequency of the qubit $\omega_0$ (see Appendix~\eqref{eq:B}). The red circles represent the second-order dynamic sweet spot at zero detuning at $\omega_d = 57.55 \, \mu\text{eV}$ ($20.72 \, \mu\text{eV}$) in the Ge (Si) DQD case in Fig.~\ref{fig:level3}(b) (Fig.~\ref{fig:level3}(d)). Other parameters are the same as those in Fig.~\ref{fig:level2}.}
\label{fig:level3}
\end{figure*}

In Fig.~\ref{fig:level3}(a), we have depicted the first derivative of the Rabi frequency $\partial_{\epsilon_{0}}\Omega$ for the hole in Ge DQD as a function of drive frequency  $\omega_d$ and detuning $\epsilon_0$, in the slightly off-resonant regime (i.e., \( \omega_d \neq \omega_0 \)). The dashed green curve represents the resonance frequency of the qubit $\omega_0$. Notably, two significant observations can be drawn from this curve.

The first observation is the manifestation of a \textit{first-order dynamic sweet spot}.
   We find that at zero detuning, the first derivative of the Rabi frequency is zero for all drive frequencies. Additionally, in the case of a non-zero detuning, there are specific drive frequencies where the first derivative of the Rabi frequency $\partial_{\epsilon_{0}}\Omega$ becomes zero. This distinctive behavior is illustrated by the yellow dashed curve, marked as \(\omega_d = \omega_0 +  \Omega_0 \partial_{\epsilon_{0}}\Omega_0 / \partial_{\epsilon_{0}}\omega_0\). The points where the first derivative of the Rabi frequency becomes zero on this curve indicate the formation of first-order dynamic sweet spots within the DQD. Crucially, the robust manifestation of the first-order dynamic sweet spot extends beyond scenarios with zero detuning, emphasizing its resilience even in the presence of detuning. It is noteworthy that as detuning increases, the values of the drive frequency \( \omega_d \) at which the first derivative is zero tend to converge toward the qubit resonance (green dashed line). This behavior is indicative of the dynamic sweet spot adapting to changes in the detuning parameter.

The second observation is the appearance of a 
\textit{second-order dynamic sweet spot} at zero detuning.
  Here, we find that the yellow dashed curve, which represents the formation of the first-order dynamic sweet spot, exhibits a minimum at zero detuning (\(\epsilon_0 = 0\)). Consequently, at this point, the second derivative of the Rabi frequency \( \partial^2_{\epsilon_{0}}\Omega \) is zero at \( \omega_d = 57.55 \, \mu\text{eV} \), indicating the presence of a second-order sweet spot, as marked by a red circle in Fig.~\ref{fig:level3}(b). This finding suggests that the system displays heightened resistance to second-order noise at this specific point. The increased robustness against second-order perturbations enhances the stability of the DQD, offering valuable insights for potential applications in quantum information processing. The emergence of the first-order dynamic sweet spot in Fig.~\ref{fig:level3}(c) and the second-order dynamic sweet spots at a drive frequency of $\omega_d = 20.72 \, \mu\text{eV}$, indicated by a red circle in Fig.~\ref{fig:level3}(d), in a Si DQD at different drive frequencies compared to the Ge DQD case is attributed to differences in tunneling coupling, Zeeman field, and SOI between Si and Ge DQDs.

\begin{figure}[t]
\centering
\includegraphics[width=1.0\linewidth]{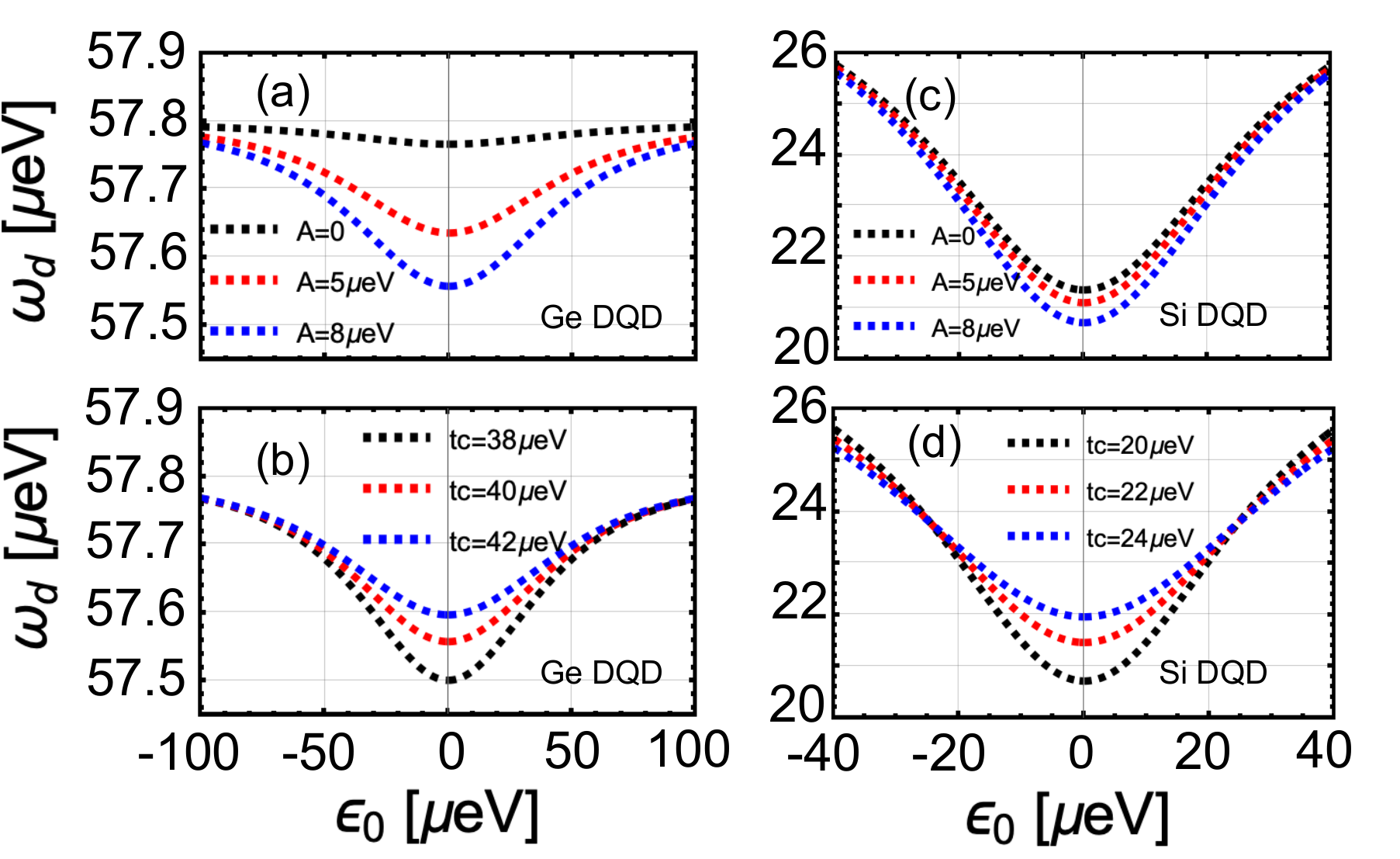} 
\caption{\label{fig:level4}
  Tunability of the first and second-order dynamic sweet spots with drive amplitude \(A\), tunneling coupling \(t_c\), and detuning $\epsilon_0$ for a Ge DQD in (a) and (b) and for a Si DQD in (c) and (d). Other parameters are the same as those in Fig.~\ref{fig:level2}.
  }
\end{figure}

In Fig.~\ref{fig:level4}(a) (Fig.~\ref{fig:level4}(c)), we have plotted the drive frequency $\omega_d$, at which $\partial_{\epsilon_{0}}\Omega=0$, as a function of detuning $\epsilon_0$ for different drive amplitudes $A$ for a Ge (Si) DQD. Our exploration involved varying drive amplitudes to analyze their impact on the system's behavior. It is observed that as the drive amplitude increases, the drive frequency associated with the first-order dynamic sweet spot extends over a broader range of drive frequencies. This effect is particularly pronounced around zero detuning and for detuned cases, the drive amplitude does not exhibit a strong effect, and all curves tend to converge towards the qubit resonance frequency $\omega_0$. Additionally, we observed that by decreasing the drive amplitude, the second-order dynamic sweet spots which are located at $\epsilon_0 = 0$ tend to approach the qubit resonance frequency $\omega_0$, further enhancing our understanding of the nuanced behavior of the system. In Fig.~\ref{fig:level4}(b) (Fig.~\ref{fig:level4}(d)), we have plotted the drive frequency $\omega_d$, at which $\partial_{\epsilon_{0}}\Omega=0$ as a function of detuning $\epsilon_0$ for different tunneling coupling $t_c$ for the Ge (Si) DQD case. An increase in tunneling coupling $t_c$ is evident in the expansion of the first-order dynamic sweet spot, covering a narrower range of drive frequencies, which contrasts with the effect of drive amplitude. The findings presented in Fig.~\ref{fig:level4} collectively suggest that both the drive amplitude and tunneling coupling play pivotal roles as control parameters influencing the characteristics of the first-order and, consequently, the second-order dynamic sweet spots. The modulation of these parameters allows for precise customization of the DQD's response to external stimuli, enhancing its versatility in quantum information processing applications. These results contribute to a deeper comprehension of the intricate interplay between system parameters and quantum dynamics within the slightly off-resonant regime of the DQD system. The concept of the dynamic sweet spot, especially in the context of Rabi oscillations, involves continuous adjustment of external field parameters during quantum computation to adapt to evolving conditions. This adaptability aims to optimize the coherence and fidelity of quantum operations, offering a more flexible and robust approach compared to a static sweet spot where parameters remain constant throughout the computation.

\begin{figure}[t]
\centering
\includegraphics[width=1\linewidth]{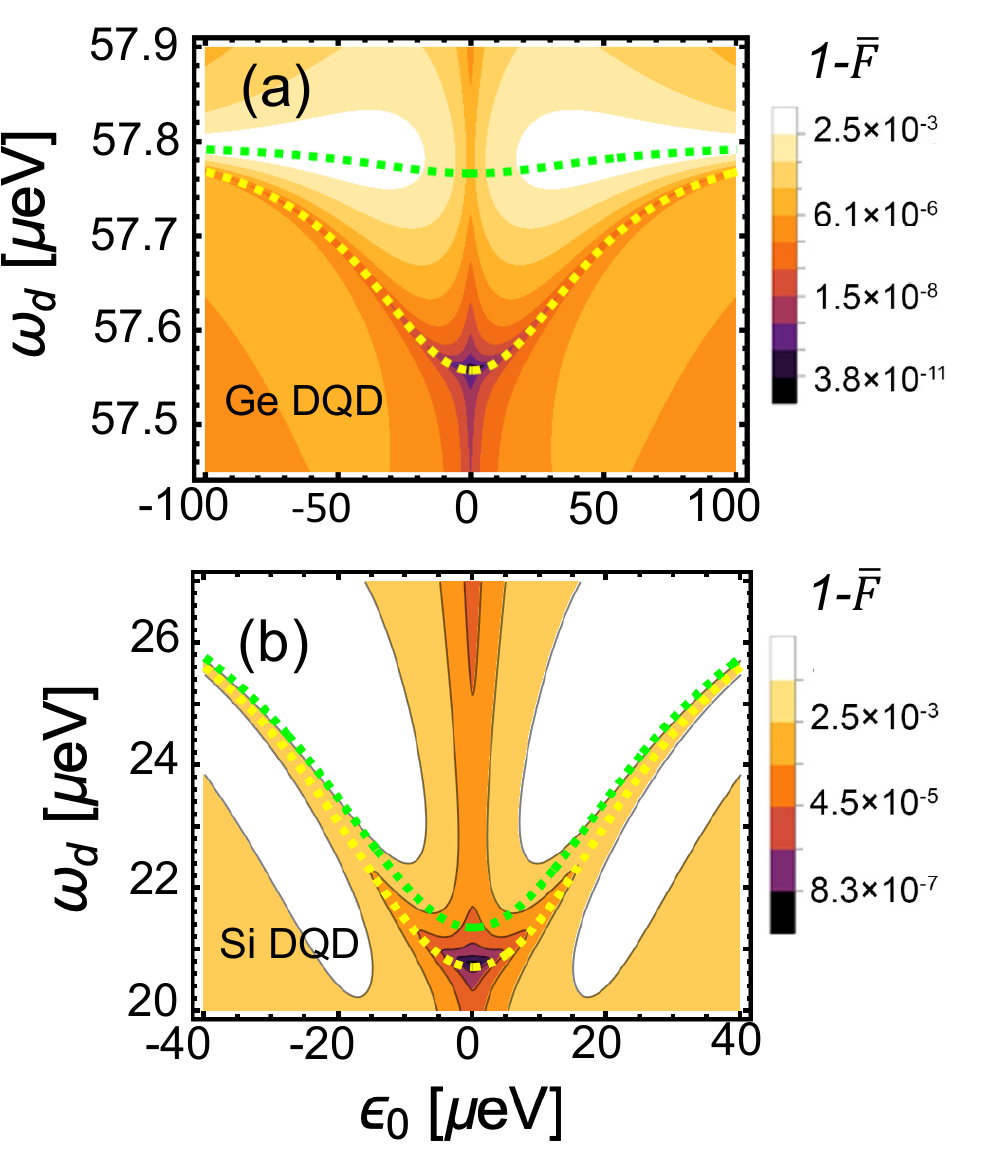} 
\caption{\label{fig:level5}
  The average gate infidelity, \(1 - \overline{F}\), of the driven qubit in the presence of static charge noise as a function of drive frequency $\omega_{d}$ and detuning $\epsilon_0$ for (a) the Ge DQD and (b) the Si DQD case. The yellow dashed line indicates the emergence of the first-order dynamic sweet spot (i.e., $\partial_{\epsilon_{0}}\Omega=0$), while the green dashed line represents the resonance frequency of the qubit $\omega_0$ (see Appendix~\eqref{eq:B}). Here, the standard deviation is \(\sigma \ = 1.5 \, \mu\text{eV}\) and other parameters are the same as those in Fig.~\ref{fig:level2}.
  }
\end{figure}

The robustness of the flopping-mode spin qubit against noise becomes apparent when directly compared with EDSR in a single QD. In this section, we evaluate the  overall performance of the qubit under various operational conditions by estimating the infidelity shown in Fig.~\ref{fig:level5}. Specifically, we analyze the average gate infidelity, \(1 - \bar{F}\), of the single-qubit system. This arises when a single QD is driven by AC fields that periodically shift the dot levels while being exposed to charge noise, considered akin to quasi-static electric noise. To calculate the gate infidelity, we use Eq.~\eqref{eq:F} where the desired qubit operation \(U_{\text{ideal}}\) is the unitary transformation representing a DQD qubit slightly off-resonance and the actual qubit operation \(U_{\text{actual}}\) is influenced by static charge noise (i.e., \( \epsilon_{0} \rightarrow \epsilon_{0} + \delta\epsilon_{0} \)), as follows,
\begin{flalign}
 \label{eq:u8}
U_{\text{ideal}} &= e^{-iH_{\text{RWA}}t} = e^{-\frac{ (i\sigma \cdot \mathbf{n}) {\Omega}t}{2}} \notag \\
&= \begin{pmatrix}
\cos(\frac{{\Omega}t}{2}) - i n_z \sin(\frac{{\Omega}t}{2}) & -i n_x \sin(\frac{{\Omega}t}{2}) \\
-i n_x \sin(\frac{{\Omega}t}{2}) & \cos(\frac{{\Omega}t}{2}) + i n_z \sin(\frac{{\Omega}t}{2})
\end{pmatrix},
\end{flalign}
where we have introduced,
\begin{flalign}
n_x &= \frac{\Omega_0(\epsilon_0)}{{\Omega}(\epsilon_0, \omega_d)}, \quad
n_z = \frac{\Delta\omega(\epsilon_0)}{{\Omega}(\epsilon_0, \omega_d)}. \label{eq:n_x_n_z} 
\end{flalign}
The actual gate operation is,
\begin{flalign}
U_{\text{actual}} &= \notag \\
&\begin{pmatrix}
    \cos(\frac{{\Omega'}t}{2}) - i n_z' \sin(\frac{{\Omega'}t}{2}) & -i n_x' \sin(\frac{{\Omega'}t}{2}) \\
    -i n_x' \sin(\frac{{\Omega'}t}{2}) & \cos(\frac{{\Omega'}t}{2}) + i n_z' \sin(\frac{{\Omega'}t}{2}) 
\end{pmatrix}, \label{eq:U_actual} &
\end{flalign}
with the definitions,
\begin{flalign}
n_x' &= \frac{\Omega_0'(\epsilon_0 + \delta\epsilon_0)}{{\Omega'}(\epsilon_0 + \delta\epsilon_0, \omega_d)}, \quad
n_z' = \frac{\Delta\omega_0'(\epsilon_0 + \delta\epsilon_0)}{{\Omega'}(\epsilon_0 + \delta\epsilon_0, \omega_d)} .\label{eq:n_x_prime_n_z_prime} &
\end{flalign}

To ensure robustness in our findings, we perform an averaging process over a Gaussian distribution for static charge noise, with a standard deviation $\sigma$. The infidelity varies with detuning $\epsilon_0$ and drive frequency $\omega_{d}$ in both Ge and Si DQD based qubits, as depicted in Fig.~\ref{fig:level5}(a) and Fig.~\ref{fig:level5}(b), respectively, at the gate time $t = \pi/\Omega$. In these figures, the green dashed line represents the qubit resonance frequency $\omega_{0}$ while the yellow curve corresponds to the frequency at which $\partial_{\epsilon_{0}}\Omega=0$, indicating the first-order dynamic sweet spot. Essentially, we compute the average gate infidelity for a slightly off-resonant case.  Under the assumption that the primary source of decoherence arises from the pure dephasing rate induced by fluctuations induced by charge noise with \(\sigma \ = 1.5 \, \mu\text{eV}\), the optimal qubit quality with fidelity higher than 99.9\% aligns with the positions of the dynamic sweet spots (dashed yellow lines) as shown in Figs.~\ref{fig:level5}(a) and~\ref{fig:level5}(b). These dynamic sweet spots, characterized by very low infidelity, are discernible not only in scenarios with zero detuning but also extend to detuned cases, as confirmed by the first derivative of the Rabi frequency curves in Fig.~\ref{fig:level3}. This observation underscores the significance of dynamic sweet spots in enhancing qubit quality, in line with our emphasis on mitigating the impact of charge noise fluctuations for improved quantum computational performance.


\begin{figure*}[ht!]
\centering
\includegraphics[width=0.8\linewidth]{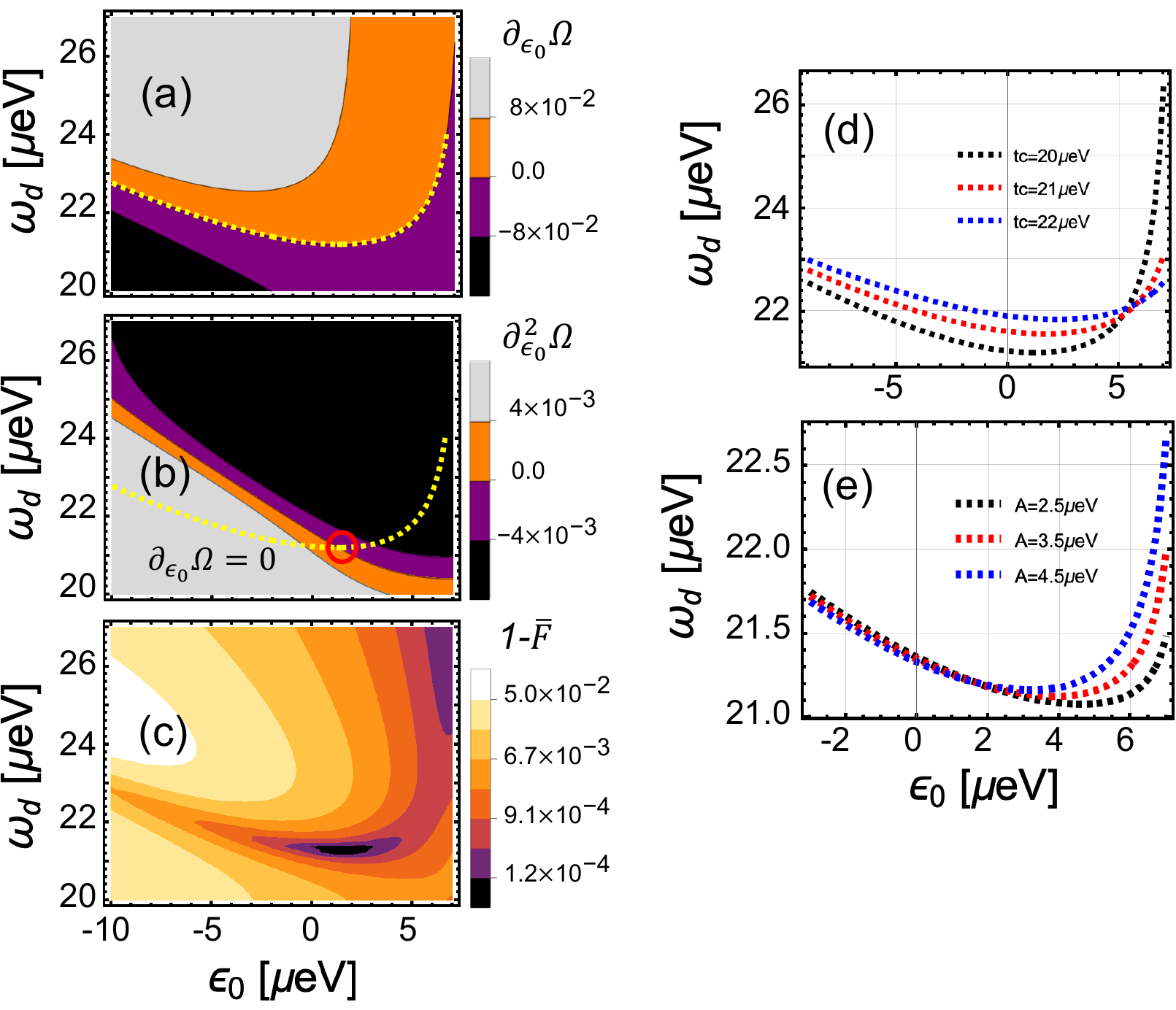} 
\caption{\label{fig:level6}
  (a) First derivative of the Rabi frequency $\partial_{\epsilon_{0}}\Omega$, (b) second derivative of the Rabi frequency $\partial^2_{\epsilon_{0}}\Omega$, and (c) average gate infidelity \(1 - \bar{F}\) for the Si DQD based qubit as a function of drive frequency $\omega_d$ and detuning $\epsilon_0$ in the presence of a longitudinal magnetic field gradient (\( b_z = 4 \, \mu\text{eV} \)). The dashed yellow lines show the first-order dynamic sweet spot. The red circle in (b) represents the formation of the second-order sweet spot, which is at \( \varepsilon_0 = 1.23 \, \mu\text{eV} \). In (d) and (e) the drive frequency $\omega_d$ is shown as a function of detuning $\epsilon_0$ for different drive amplitude A and tunneling coupling $t_c$, respectively. Other parameters are the same as those in Fig.~\ref{fig:level2}.
  }
\end{figure*}

In Figs.~\ref{fig:level3} and~\ref{fig:level4}, it is evident that the second-order dynamic sweet spot for both the Ge and Si DQD case (in the absence of a longitudinal magnetic field gradient, \( b_z \)) is exclusively observed at zero detuning. However, the introduction of a longitudinal magnetic field gradient \( b_z \) in a Si DQD leads to the induction of distinct spin resonance frequencies depending on the electron position between the two dots, resulting in a unique scenario as depicted in Fig.~\ref{fig:level6}. In these figures, the yellow dashed curves correspond to the frequency at which \( \partial_{\epsilon_{0}}\Omega=0 \). Firstly, the first-order dynamic sweet spot is no longer symmetric with respect to detuning \( \epsilon_{0} \), as observed in \( \partial_{\epsilon_{0}}\Omega \) curve in Fig.~\ref{fig:level6}(a). Secondly, in the \(\partial^2_{\epsilon_{0}}\Omega\) curve shown in Fig.~\ref{fig:level6}(b), the yellow dashed line indicating the first-order dynamic sweet spot intersects the zero line of the second derivative, resulting in a second-order dynamic sweet spot at \(\varepsilon_0 = 1.23 \, \mu\text{eV}\) for \(t_c = 20 \, \mu\text{eV}\) and \(A = 8 \, \mu\text{eV}\), as marked by the red circle.
 This is because the longitudinal magnetic field gradient \( b_z \) in Si DQDs contributes to the diagonal term of the Hamiltonian in Eq.~\eqref{eq:Ham} and couples with the detuning leading to a different spin resonance frequency depending on the electron position between two dots, unlike the natural SOI terms in Ge DQDs (\( \eta \) and \( \chi \)), which do not appear in the diagonal term of the Hamiltonian in Eq.~\eqref{eq:Ham}. Our observations in Figs.~\ref{fig:level6}(a) and~\ref{fig:level6}(b) can be further confirmed by the infidelity curve in Fig.~\ref{fig:level6}(c) which is no longer symmetric with respect to detuning and shows the best fidelity at detuned case (\( \varepsilon_0 = 1.23 \, \mu\text{eV} \)). In Figs.~\ref{fig:level6}(d) and~\ref{fig:level6}(e), we have shown the drive frequency \( \omega_d \) corresponding to the first-order dynamic sweet spot as a function of detuning \( \epsilon_0 \) for different tunneling coupling \( t_c \) and drive amplitude \( A \), respectively. It is seen that the position of the second-order dynamic sweet spot, corresponding to the minimum of \( \omega_d \) curves, is tunable with both the tunneling coupling and drive amplitude values. This tunability underscores the dynamic nature of the system, offering valuable insights into how specific parameters influence the occurrence of optimal conditions for resonance. This configuration protects against charge noise up to the second-order in detuned cases for the qubit which is tunable by drive and DQD parameters. Similar behavior, observed in the presence of longitudinal magnetic fields gradient \( b_z \) for the static sweet spot in Si DQD \cite{monica}, differs in our case as the position of the second-order dynamic sweet spot is tunable by drive parameters (\( A \) and \( \omega_d \)).

\begin{figure}[t]
\centering
\includegraphics[width=1.0\linewidth]{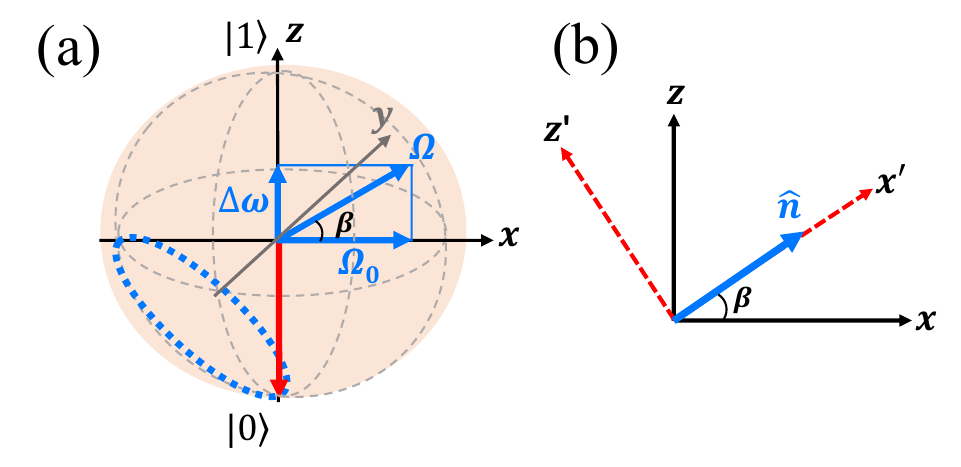} 
\caption{\label{fig:level7}
  (a) Nutation of a spin around the Bloch sphere occurs in the rotating frame, initialized in the \( |0\rangle \) state, for a slightly off-resonant regime (i.e., \( \Delta\omega \neq 0 \): \(\Omega = \sqrt{\Omega_0^2 + \Delta\omega^2}\)). 
  (b) The rotation angle \(\boldsymbol{\hat{n}}\) with its projection onto the \(xz\) plane.
 We establish a new coordinate system with axes \( x^\prime \) and \( z^\prime \), such that \( n^\prime_z = 0 \), by rotating the \( xz \) plane around the \( y \) axis by an angle \( \beta \). Here, \( \Delta\omega \) is the detuning between driving and resonance frequencies and $\Omega_0$ and $\Omega$ are the Rabi frequencies under on-resonant and off-resonant driving, respectively. 
}
\end{figure}

After establishing optimal conditions for our high-fidelity spin qubit, we aim to create a gate aligning the Rabi vector with the \( x\)-axis via frame rotation. Since z-rotations are easily implemented, we focus on \( x\)-rotations. To form an \( X\) gate from the unitary matrix for \(xz\)-plane rotations in Eq.~\eqref{eq:u8}, we rotate the frame to align the \( \hat{n} \) vector with the \( x\)-axis. This alignment is achieved by a rotation angle \( \beta = \arctan(n_z/n_x) = \arctan(\Delta\omega/\Omega) \), where \((n_z, n_x)\) are the components of \( \hat{n} \). After alignment, we perform a \( \pi \) rotation around the new \(x\)-axis to obtain the \( X\) gate. The modified unitary matrix for the \( X\) gate is \( X = R_x(\beta) U_{\text{real}} R_x(\pi) \), where \( R_x(\beta)=e^{i \beta \sigma_x/2} \) rotates around the \( x\)-axis by angle \( \beta \), and \( U_{\text{real}} \) is the original matrix for \( xz\)-plane rotations in Eq.~\eqref{eq:u8}. Subsequently, another \( \pi \) rotation around this new \( x\)-axis yields the \( X\) gate. In the off-resonant regime, the rotation axis is close to one of the poles of the Bloch sphere. As a result, the trajectory traced by the state on the Bloch sphere remains near the south pole, indicating a small amplitude, as illustrated in Fig.~\ref{fig:level7}(a). Conversely, for the \( X\) gate applied within the rotated frame, depicted in Fig.~\ref{fig:level7}(b), the rotation axis aligns with the equator of the Bloch sphere. Consequently, the trajectory of the state extends to the north pole, indicative of a significantly larger amplitude.

\section{Experimental feasibility}\label{sec:stimtmm}
A hole spin qubit based on a planar and hut wire DQD in germanium has been synthesized, as reported in \cite{jirovec,jirovec2,zhang,liu,wang}. Furthermore, it is noteworthy that flopping-mode electric dipole spin resonance, involving the electrical driving of an electron in a Si/SiGe double quantum dot in the presence of a substantial magnetic field gradient, has been successfully demonstrated \cite{croot,Hu}. In the context of our investigation into a DQD hole qubit implemented in germanium, the careful manipulation of hole transfer from the left to the right QD reveals subtle variations in the g-factor between these dots. Although the pronounced modulation of the g-factor may be a primary contributor to decoherence phenomena, strategic parameter tuning presents an opportunity to engineer quantum dots with uniform g-factors. This intricate optimization process, focusing on the electric field landscape, is anticipated to yield a controlled qubit Zeeman splitting. Such control holds the potential to create conditions conducive to heightened coherence by minimizing the local slope of the system. It is noteworthy that the impact of g-tensor modulation and g-factor differences becomes particularly significant for Rabi frequency only when the applied magnetic field direction is non-perpendicular to the sample \cite{jirovec,miz,liu} which is not the case in our study.  

\section{Conclusion}\label{sec:stimconc}

In summary, we explore the impact of charge noise on a driven planar DQD spin qubit, specifically focusing on the flopping-mode with SOI. Our findings demonstrate that driving the qubit slightly off-resonantly effectively mitigates charge noise influence, revealing a dynamic sweet spot manifestation not only at zero detuning but also in detuned cases. This modulation significantly improves quantum gate fidelity within specific ranges of drive parameters and detuning. Additionally, our study reveals the potential for inducing a second-order sweet spot in the proposed qubits, tunable by drive, SOI, and DQD parameters. Understanding the importance of driving qubits off-resonantly is crucial for developing high-coherence planar DQD spin qubits for both electrons in silicon and holes in germanium, enhancing their performance in quantum information processing.

\begin{acknowledgments}
  We acknowledge financial support from the IGNITE project of European Union’s Horizon Europe Framework Programme under grant agreement No.~101069515.
\end{acknowledgments}

\appendix
\section{Time-dependent Schrieffer-Wolff transformation (TDSWT)}\label{eq:A}
The time-dependent Schrieffer-Wolff transformation (TDSWT) \cite{Romi} is a perturbative method used to derive an effective block diagonal 2 × 2 time-dependent Hamiltonian $\tilde{H}(t)$ from a dense Hamiltonian $H(t)$ such as Eq.~\eqref{eq:Ham}. This process involves applying the unitary transformation $U(t) = e^{-S(t)}$, where $S(t)$ is the generator of the transformation,
\begin{equation}
    \tilde{\psi}(t) = e^{-S(t)} \psi(t), \tag{A1}
\end{equation}
and, using the time-dependent Schr\"{o}dinger equation,
\begin{equation}
    -i\hbar \frac{\partial \psi(t)}{\partial t} + H(t)\psi(t) = 0, \tag{A2}
\end{equation}
leading to the transformed Hamiltonian
\begin{equation}
 \label{eq:Ht}
    \tilde{H}(t) = e^{-S(t)} H(t) e^{S(t)} + i\hbar \frac{\partial e^{S(t)}}{\partial t}. \tag{A3}
\end{equation}
Here, $S(t)$ represents a block off-diagonal matrix. It is noteworthy that due to the weak inter-subspace coupling, the unitary transformation $U(t)$ is approximately unity, making $S(t)$ small and expressible as a power series in the perturbation. This setup allows for the application of a power-series expansion, which can be further simplified to yield
\begin{equation}
 \label{eq:Hbar}
    \tilde{H}(t) = \sum_{j=0}^{\infty} \frac{1}{j!}[H(t),S]^{(j)} - i \sum_{j=0}^{\infty} \frac{1}{(j+1)!}[\dot{S}(t),S]^{(j)}. \tag{A4}
\end{equation}
At this stage, we determine $S(t)$ by assuming that $\tilde{H}(t)_{\text{off-diagonal}} = 0$. Currently, no approximations have been introduced. The strategy employed to address Eq.~\eqref{eq:Ht} and achieve $\tilde{H}(t)_{\text{off-diagonal}} = 0$ involves a power-series expansion of the small parameters (time-dependent detuning) within the matrix $S(t)$,
\[ S(t) = S(t)^{(1)} + S(t)^{(2)} + S(t)^{(3)} + \ldots \] 
where $S(t)^{(n)}$ represents the nth order of the power series.
Note that the time-dependent Hamiltonian in Eq.~\eqref{eq:Ham} can be written as \(H(t) = H_0 + H'(t)\), where the perturbation is decomposed into a block-diagonal part, \(H_1\), and a block-off-diagonal part, \(H_2\), \(H'(t) = H_1(t) + H_2(t)\) as follows,
\begin{equation}
\hspace{-5mm}
H_0 =
\begin{pmatrix}
 K_{+} & 0 & 0 & 0 \\
 0 & K_{-} & 0 & 0 \\
 0 & 0 & -K_{-} & 0 \\
 0 & 0 & 0 & -K_{+} \\
\end{pmatrix}, \tag{A5}
\end{equation}
\begin{widetext}
\begin{equation}
\begingroup 
\setlength\arraycolsep{-12pt}
H_1(t) = \begin{pmatrix}
 K_{+} + \frac{1}{2}\left(A_{c} \cos(\omega_{d}t) + b_{zc}\right) & \rho_{c} & 0 & 0 \\
 \rho_{c} & K_{-} + \frac{1}{2}\left(A_{c} \cos(\omega_{d}t) - b_{zc}\right) & 0 & 0 \\
 0 & 0 & -K_{-} - \frac{1}{2}\left(A_{c} \cos(\omega_{d}t) + b_{zc}\right) & -\rho_{c} \\
 0 & 0 & -\rho_{c} & -K_{+} - \frac{1}{2}\left(A_{c} \cos(\omega_{d}t) - b_{zc}\right) \\
\end{pmatrix}, \tag{A6}
\endgroup
\end{equation}
\begin{equation}
H_2(t) = \begin{pmatrix}
 0 & 0 & -\frac{1}{2}\left(b_{zs} + A_{s} \cos(\omega_{d}t)\right) & -\eta - \rho_{s} \\
 0 & 0 & \eta - \rho_{s} & \frac{1}{2}\left(b_{zs} - A_{s} \cos(\omega_{d}t)\right) \\
 -\frac{1}{2}\left(b_{zs} + A_{s} \cos(\omega_{d}t)\right) & \eta - \rho_{s} & 0 & 0 \\
 -\eta - \rho_{s} & \frac{1}{2}\left(b_{zs} - A_{s} \cos(\omega_{d}t)\right) & 0 & 0 \\
\end{pmatrix}. \tag{A7}
\end{equation}
\end{widetext}
The TDSWT up to the second-order is given by
\begin{align}
\label{eq:H1}
    \tilde{H}(t) = \tilde{H}_0 + \tilde{H}_1(t) + \tilde{H}_2(t). \tag{A8}
\end{align}
Here, the expansions in $\tilde{H}(t)$ can be solved from Eq.~\eqref{eq:Hbar} as
\begin{align}
    \tilde{H}_0 & = H_0, \tag{Aa9}\\
    \tilde{H}_1 (t)& = H_1(t), \tag{Ab9}\\
    \tilde{H}_2(t) & = \frac{1}{2\hbar} [H_2(t), S(t)^{(1)}].\tag{Ac9}
\end{align}
Only $S(t)^{(1)}$ needs to be solved for, a task accomplished by applying the \(\tilde{H}(t)_{\text{off-diagonal}}\) condition, yielding \( [ H_0, S(t)^{(1)} ]  = -H_2(t)\). Finally, the Hamiltonian in  Eq.~\eqref{eq:Ham} can be approximated using TDSWT in Eq.~\eqref{eq:H1} to a block diagonal form, encompassing bonding and antibonding states as presented in  Eq.~\eqref{eq:TDSWT}. Hence, \(H_{\text{Bonding}}\) for Ge and Si DQDs is given by
\begin{widetext}
\begin{equation}
\label{eq:H2}
\hspace{-5mm}
H_{\text{Ge-Bonding}}=
\begin{pmatrix}
    -\frac{2\Gamma_{+}(2K_{-}^2+(\eta+\chi_s)^2)+K_{-} \cos(\omega_d t)(2A_c\Gamma_{+}+A_s^2 \cos(\omega_d t))}{4K_{-} \Gamma_{+}} & \chi_c+\frac{A_s \left(6K_{-} K_{+} \chi_s+K_{-}^2 (-\eta+\chi_s)+K_{+}^2 (\eta+\chi_s)\right)\cos(\omega_d t)}{8K_{-} K_{+} \Gamma_{+}} \\
    \chi_c+\frac{A_s \left(6K_{-} K_{+} \chi_s+K_{-}^2 (-\eta+\chi_s)+K_{+}^2 (\eta+\chi_s)\right)\cos(\omega_d t)}{8K_{-} K_{+} \Gamma_{+}} & -\frac{2\Gamma_{+}(2K_{+}^2+(\eta-\chi_s)^2)+K_{+} \cos(\omega_d t)(2A_c\Gamma_{+}+A_s^2 \cos(\omega_d t))}{4K_{+} \Gamma_{+}}
\end{pmatrix},
\tag{A10}
\end{equation}
\begin{equation}
\label{eq:H3}
\hspace{-5mm}
\begingroup 
\setlength\arraycolsep{-4pt}
H_{\text{Si-Bonding}}=
\begin{pmatrix}
-\frac{b_{xs}^2 \Gamma_{+} + \delta_{-} + 2 K_{-} \cos( \omega_d t) (2 A_s b_{zs} + 2 A_c \Gamma_{+} + A_s^2 \cos( \omega_d t))}{8 K_{-} \Gamma_{+}} & -\frac{\Gamma_{+} (b_{xs} b_{zs} \Gamma_{-} + 8 b_{xc} K_{-} K_{+}) + A_s b_{xs} \Delta \cos( \omega_d t)}{16 K_{-} K_{+} \Gamma_{+}} \\
-\frac{\Gamma_{+} (b_{xs} b_{zs} \Gamma_{-} + 8 b_{xc} K_{-} K_{+}) + A_s b_{xs} \Delta \cos( \omega_d t)}{16 K_{-} K_{+} \Gamma_{+}} & -\frac{ b_{xs}^2 \Gamma_{+} + \delta_{+} + 2 K_{+} \cos( \omega_d t) (-2 A_s b_{zs} + 2 A_c \Gamma_{+} + A_s^2 \cos( \omega_d t))}{8 K_{+} \Gamma_{+}}
\end{pmatrix},
\tag{A11}
\endgroup
\end{equation}
\end{widetext}
in which $\Delta=(K_{-}^2+6K_{-}K_{+}+K_{+}^2)$, $\delta_{\pm}=2 K_{\pm} (b_{zs}^2 {\mp} 2( b_{zc} {\mp} 2 K_{\pm})\Gamma_{+})$, and $\Gamma_\pm=(K_{-}\pm K_{+})$.

\section{Rotating wave approximation (RWA)}\label{eq:B}
In this section, we illustrate the derivation of the Rabi frequency for bonding states in Ge and Si DQDs using the RWA.
For HHs in Ge DQD we can write the $H_{\text{Ge-Bonding}}$ in Eq.~\eqref{eq:H2} as 
\begin{equation}
\label{eq:H4}
H_{\text{Ge-Bonding}} = H_{0} + H_{1} \cos(\omega_d t) + H_{2} \cos^2(\omega_d t),
\tag{B1}
\end{equation}
where 
\begin{equation}
\label{eq:b2}
\begingroup 
\setlength\arraycolsep{-6pt}
\hspace{-5mm}
H_0 =
\begin{pmatrix} 
-\frac{2K_{-}^2+(\eta+\chi_s)^2}{2K_{-}} & \chi_c \\[1.5pt]
\chi_c & -\frac{2K_{+}^2+(\eta-\chi_s)^2}{2K_{+}}
\end{pmatrix},
\endgroup\tag{B2}
\end{equation}
\begin{widetext}
\begin{equation}
\begingroup 
\setlength\arraycolsep{-15pt}
H_1 = \begin{pmatrix} 
-\frac{A_c}{2} & \frac{A_s(6K_{-} K_{+} \chi_s+K_{-}^2 (-\eta+\chi_s)+K_{+}^2 (\eta+\chi_s))}{8K_{-} K_{+} \Gamma_{+}}  \\[1.5pt]
\frac{A_s(6K_{-} K_{+} \chi_s+K_{-}^2 (-\eta+\chi_s)+K_{+}^2 (\eta+\chi_s))}{8K_{-} K_{+}\Gamma_{+} } & -\frac{A_c}{2} 
\end{pmatrix}
\endgroup, \quad
H_2 = \begin{pmatrix} 
\frac{A_s^2}{4 \Gamma_{+}} & 0  \\[1.5pt] 
0 & \frac{A_s^2}{4 \Gamma_{+}} 
\end{pmatrix}. \tag{B3} 
\end{equation}
\end{widetext}
Upon diagonalizing  \(H_{0}\) in Eq.~\eqref{eq:b2}, the new eigenvalues of the system read as
\begin{multline*}
\label{eq:b4}
E_{\pm} = \dfrac{-\Gamma_{+}}{2} - \frac{(\eta - \chi_s)^2}{4 K_{+}} - \frac{(\eta + \chi_s)^2}{4 K_{-}} \\
\pm \dfrac{1}{2}\sqrt{4 \chi_c^2 + \frac{1}{4} \left(2 \Gamma_{-} - \frac{(\eta - \chi_s)^2}{K_{+}} + \frac{(\eta + \chi_s)^2}{K_{-}}\right)^2}.
\tag{B4}
\end{multline*}
The energy level splitting of the qubit is given by 
\begin{multline*}
\tag{B5}
\omega_0 = E_{+} - E_{-} = \\\sqrt{4 \chi_c^2 + \frac{1}{4} \left(2 \Gamma_{-} - \frac{(\eta - \chi_s)^2}{K_{+}} + \frac{(\eta + \chi_s)^2}{K_{-}}\right)^2}.
\end{multline*}
If we express the $H_{\text{Si-Bonding}}$ in Eq.~\eqref{eq:H3} in the same manner as in Eq.~\eqref{eq:H4}, we can calculate the eigenenergies and energy level splitting of the electron in Si DQD as 
\begin{widetext}
\begin{center}
\begin{multline*}
\tag{B6}
E_{\pm} = \frac{1}{2} \left( -b_{zc} + \frac{(b_{xs}^2 - 8 K_{-} K_{+}) \Gamma_{-}}{8 K_{-} K_{+}} \pm \sqrt{\frac{(8 b_{xc} K_{-} K_{+} + b_{xs} b_{zs} \Gamma_{-})^2}{64 K_{-}^2 K_{+}^2} + \left(-b_{zc} + \frac{(b_{xs}^2 - 8 K_{-} K_{+}) \Gamma_{-}}{8 K_{-} K_{+}}\right)^2} \right),
\end{multline*}
\end{center}
\begin{equation*}
\tag{B7}
\begin{gathered}
\omega_0 = \sqrt{\frac{(8 b_{xc} K_{-} K_{+} + b_{xs} b_{zs} \Gamma_{-})^2}{64 K_{-}^2 K_{+}^2} + \left(-b_{zc} + \frac{(b_{xs}^2 - 8 K_{-} K_{+}) \Gamma_{-}}{8 K_{-} K_{+}}\right)^2}.
\end{gathered}
\end{equation*}
\end{widetext}
Now, in the rotated frame defined by Eq.~\eqref{eq:RWA1} with $U = \text{diag}[e^{-i\omega_0 t/2}, e^{i\omega_0 t/2}]$, corresponding to resonant driving, the resulting Rabi frequency $\Omega_0$, obtained by employing RWA, for Ge and Si DQDs is given by
\begin{widetext}
\begin{equation}
\label{eq:b8}
\tag{B8}
\Omega_{0,{\text{GeDQD}}} = \frac{A_s (6K_{-} K_{+} \chi_s + K_{-}^2 (-\eta + \chi_s) + K_{+}^2 (\eta + \chi_s))}{8K_{-} K_{+} \Gamma_{+}\sqrt{1 + \frac{4\chi_c^2}{\left(\frac{-2(2K_{-}^2 + (\eta + \chi_s)^2)}{4K_{-}} + \frac{2(2K_{+}^2 + (\eta - \chi_s)^2)}{4K_{+}}\right)^2}}},
\end{equation}
\begin{equation}
\label{eq:b9}
\tag{B9}
\begin{gathered}
\Omega_{0,{\text{SiDQD}}} = 
\frac{A_{s}( -64 b_{xc} b_{zs} K_{-}^2 K_{+}^2 - b_{xs}^3 \Gamma_{-} \Delta + 8 b_{xs} K_{-} K_{+} ( (b_{zc} + \Gamma_{-}) \Delta - b_{zs}^2 \Gamma_{-}))} {128 K_{-}^2 K_{+}^2 \Gamma_{+}} \times
( -b_{zc} + 
\frac{\Gamma_{-}(b_{xs}^2 - 8 K_{-} K_{+})}{8 K_{-} K_{+}})^{-1} \\[10pt]
\times \lbrace- \frac{( -8 K_{-} K_{+}( 8 K_{-} (-b_{xc}^2 + b_{zc} + \Gamma_{-} ) K_{+} - b_{xs}^2 \Gamma_{-})}{8 K_{-} K_{+} 
( 8 K_{-} (b_{zc} + \Gamma_{-} ) K_{+} - b_{xs}^2 \Gamma_{-} )}
 +\frac{ 16 b_{xc} b_{xs} b_{zs} K_{-} K_{+} \Gamma_{-} + b_{xs}^2 b_{zs}^2 \Gamma_{-}^2)}{8 K_{-} K_{+} ( 8 K_{-} (b_{zc} + \Gamma_{-} ) K_{+} - b_{xs}^2 \Gamma_{-} )}\rbrace^{-1/2}.
\end{gathered}
\end{equation}
\end{widetext}
 Note that to drive the qubit slightly off-resonantly, i.e., \(\omega_{d}\neq\omega_{0}\),
 with the unitary operator $U = \text{diag}[e^{-i\omega_d t/2}, e^{i\omega_d t/2}]$ we can derive the Rabi frequency $\Omega$ by substituting Eqs.~\eqref{eq:b8} and \eqref{eq:b9} into Eq.~\eqref{eq:R} for Ge and Si DQDs, respectively.  


\bibliography{refs.bib}


\end{document}